\documentstyle[graphics,twocolumn]{mn}

\newcommand{\mincir}{\raise
-2.truept\hbox{\rlap{\hbox{$\sim$}}\raise5.truept 
\hbox{$<$}\ }}
\newcommand{\magcir}{\raise
-2.truept\hbox{\rlap{\hbox{$\sim$}}\raise5.truept
\hbox{$>$}\ }}
\newcommand{\minmag}{\raise-2.truept\hbox{\rlap{\hbox{$<$}}\raise
6.truept\hbox
{$>$}\ }}

\newcommand{\lya}{Lyman-$\alpha$~}

\newcommand{\be}{\begin{equation}}
\newcommand{\ee}{\end{equation}}
\newcommand{\ba}{\begin{eqnarray}}
\newcommand{\ea}{\end{eqnarray}}
\newcommand{\brr}{\begin{array}}
 
\newcommand{\err}{\end{array}}
\newcommand{\bc}{\begin{center}}
\newcommand{\ec}{\end{center}}

\newcommand{\vel}{\,{\rm km\,s^{-1}}}
\newcommand{\et}{{\it et al.~}}

\newif\ifAMStwofonts
\AMStwofontstrue

\DeclareMathAlphabet{\mathsc}{OT1}{cmr}{m}{sc}
\def\testbx{bx}%
\DeclareRobustCommand{\ion}[2]{%
\relax\ifmmode
\ifx\testbx\f@series
{\mathbf{#1\,\mathsc{#2}}}\else
{\mathrm{#1\,\mathsc{#2}}}\fi
\else\textup{#1\,{\mdseries\textsc{#2}}}%
\fi}

\title[The bispectrum of the \lya forest]{The bispectrum of the \lya
forest at $z\sim 2-2.4$ from a Large sample of UVES QSO 
Absorption Spectra (LUQAS)
\thanks{Based on data taken from the ESO archive obtained with UVES at VLT,
Paranal, Chile.}}
\author[M. Viel {\it et al.}]
{M. Viel $^{1}$, S. Matarrese $^{2,3}$, A. Heavens $^{4}$,
M.G. Haehnelt $^{1}$, T.-S. Kim $^{1}$, \newauthor  V. Springel
$^{5}$ \& L. Hernquist $^{6}$ \\ 
$^1$ Institute of Astronomy, Madingley Road, Cambridge CB3 0HA\\
$^2$ Dipartimento di Fisica `Galileo Galilei', via Marzolo 8, I-35131 Padova, 
Italy \\
$^3$ INFN, Sezione di Padova, via Marzolo 8, I-35131 Padova, Italy \\
$^4$ Institute for Astronomy, University of Edinburgh, Blackford Hill, Edinburgh EH9 3HJ\\
$^5$ Max-Planck-Institut f\"ur Astrophysik, Karl Schwarzschild-str. 1,
Garching bei M\"unchen, Germany\\
$^6$ Harvard-Smithsonian Center for Astrohpysics, 60 Garden Street,
Cambridge, MA 02198, USA\\}

\begin{document}

\maketitle
\begin{abstract}
We present a determination of the bispectrum of the flux in the 
\lya forest of QSO absorption spectra obtained from LUQAS which 
consists of spectra observed  with the high resolution Echelle 
spectrograph UVES.  Typical errors on the observed bispectrum 
as obtained from a jack-knife estimator are $\sim 50$ \%. For 
wavenumbers in the range $0.03 \,{\rm s/km } <k < 0.1 \,{\rm s/km }$ 
the observed bispectrum agrees within the errors with that of the 
synthetic absorption spectra obtained from numerical hydro-simulations
of a $\Lambda$CDM model  with and without feedback from star
formation.  Including galactic
feedback changes the bispectrum by less than 10\%. 
At smaller wavenumbers the associated metal absorption lines 
contribute about 50\% to the bispectrum and the observed bispectrum 
exceeds that of the simulations. At wavenumbers $k< 0.03 \,{\rm s/km}$ 
second-order perturbation theory applied to the flux spectrum
gives a reasonable (errors smaller  than 30\%) approximation to the
bispectra of observed and simulated absorption spectra. The bispectrum 
of the observed absorption spectra also 
agrees, within the errors, with that of a randomized set of absorption 
spectra  where a random shift in wavelength has been added to  
absorption lines identified with VPFIT. This suggests that for a sample of 
the size presented here, the errors on the bispectrum are too 
large to discriminate between models with very different  3D 
distribution of \lya absorption. If it were possible to substantially 
reduce these errors for larger samples of absorption spectra, the bispectrum 
might become an important statistical tool for probing the growth of 
gravitational structure in the Universe at redshift $z \magcir 2$.  
\end{abstract}


\section{Introduction}

The \lya forest in the absorption spectra of high-redshift QSOs has
been recognized as an important tool for studying the underlying matter
distribution (see Rauch 1998 and Weinberg et al. 1999 for excellent
reviews). In a previous paper (Kim et al. 2003, K03) an analysis of the
flux power-spectrum of a large sample of UVES QSO absorption spectra
(LUQAS) has been presented. Here we investigate the bispectrum of the
flux in a subset of these absorption spectra.  

The bispectrum is the Fourier
transform of the three-point correlation function. As the flux in the \lya forest
is a sensitive probe of the matter distribution it should probe the
topology of the 3D matter distribution in more detail than the
two-point correlation function or the power-spectrum.  It also can be
used as a complementary tool to determine cosmological parameters
(e.g. Fry 1994; Verde et al. 2002) and maybe also the physical state of
the IGM. Gravitational growth induces correlations between large scale
modes and small scale power  which can be probed by the bispectrum. 
Zaldarriaga et al. (2001) pointed out that these correlations 
may be used to discriminate between fluctuations due to  
large-scale structure in the matter distribution  and those produced 
by non-gravitational processes such as fluctuations in the continuum 
emission of the quasar. Mandelbaum et al. (2003) showed 
that with the SDSS  (Sloan Digital Sky
Survey) QSO sample it should be possible to  use higher order 
statistics such as the bispectrum to determine  amplitude, 
slope and curvature of the slope of  the matter power spectrum  
with an  accuracy of a few percent provided systematic errors 
are under control. 

The plan of the paper is as follows. In Section \ref{data} 
we briefly describe our sample.  In Section \ref{method} we define the
flux bispectrum and  compare the data with hydro-dynamical
simulations, an analytical prediction and randomized spectra.
Section \ref{conclu} contains a summary of our results.

\section{The Data}
\label{data}
The LUQAS sample consists of 27 spectra taken with the Ultra-Violet Echelle
Spectrograph (UVES) on VLT (Paranal, Chile) over the period 1999--2002.
The spectra were drawn from the ESO archive and are publicly
available to the ESO community.  The  total redshift path of the
sample is $\Delta z = 13.75$. The median redshift
of the sample is $ \langle z \rangle \, =2.25$ and the number of 
spectra covering the median redshift is 17. The analysis of the
bispectrum has been restrained to the redshift range $2<z<2.4$, 
where approximately 15 QSOs contribute.
The reason for this is the strong dependence of the bispectrum on the
mean flux. The typical S/N is $\sim 50$. 

The spectra were reduced with the ECHELLE/UVES environment of the
software package MIDAS. Lists of HI absorption lines and associated
metal absorption lines identified and fitted with VPFIT (Carswell et
al.: http://www.ast.cam.ac.uk/$\sim$rfc/vpfit.html) were available for
8 and 13 QSOs, respectively. Five out of the eight QSOs with lists of
HI absorption are in the redshift range for which we calculated the
bispectrum of the flux.  We have removed four damped/sub-damped \lya
systems before computing the bispectrum.  For a more complete
description of the sample and for the analysis of the flux
power-spectrum and its evolution with redshift we refer to K03.

\section{The bispectrum of the \lya forest}
\label{method}
\subsection{Method}

\begin{figure}
\center\resizebox{.48\textwidth}{!}{\includegraphics{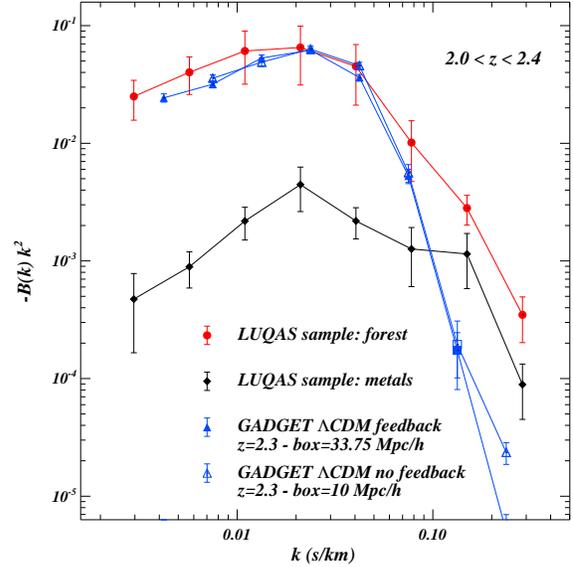}}
\caption{{\protect\footnotesize {
Filled circles show the one-dimensional bispectrum of the \lya forest
flux of the LUQAS sample in the redshift range $2<z<2.4$. The corresponding 
bispectrum of the  associated metal absorption is shown 
by filled diamonds.  Also shown is  the  flux bispectrum of synthetic
spectra computed from 
hydro-dynamical simulations of a  $\Lambda$CDM model at $z=2.3$ 
(see Section \ref{hydro}).  Filled triangles are for a simulation 
with a box size of 33.75 $h^{-1}$ Mpc which include feedback 
from galactic winds.  
Empty triangles are for a simulation with  box size of 
10  $h^{-1}$ Mpc without galactic feedback. Squares 
indicate positive values of $B(k)$. Errors are obtained with a jack-knife 
estimate.}}}
\label{fig1}
\end{figure}

We  consider the quantity  $\delta_F= (F- \langle F \rangle)/
\langle F \rangle$, where  $F$  is the continuum fitted   
transmitted flux and  $\langle F \rangle$ the average flux 
for each QSO (Hui et al. 2001, Croft et al. 2002).  
Note that $\delta_F$ corresponds to the flux estimator $F_2$ of K03.   

We follow the definition of Matarrese, Verde \& Heavens (1997) and
Verde et al. (1998) for the bispectrum.  
We use the real part of the three point 
function in $k-$space $, D_F= {\rm Re}(\delta_F (k_1)\, \delta_F (k_2) \, 
\delta_F (k_3))$, for closed triangles $ k_1 + k_2 + k_3 = 0$.  
$\delta_F(k)$ is the Fourier transform of $\delta F$. 
$D_F$ is related to the {\it bispectrum of the flux} 
$B_F(k_1,k_2,k_3)$  
\be 
\langle D_F \rangle = 
2\,\pi \,B_F(k_1, k_2, k_3)\,\delta^D (k_1 + k_2 + k_3) \; .  
\label{eq1}
\ee 
$\delta^D(k)$ is the one-dimensional Dirac delta function and 
$\langle \cdot \rangle$ indicates the ensemble average. We compute the
one-dimensional bispectrum.  Our triangles are thus degenerate and we 
choose the configuration for which $k_1=k_2$ and $k_3 =
-2\,k_1$. 
In the following we will always show the flux bispectrum as a function
of the wavenumber $k=k_1$. 

\subsection{The observed bispectrum}

The filled cirles in Fig.~\ref{fig1} show the one-dimensional bispectrum
of the \lya forest flux of the LUQAS sample in the redshift
range $2\le z \le 2.4$ plotted in dimensionless units (Table 1).  
The bin size was chosen such that visual comparison of different
bispectra in the plots is not
hindered by too large bin-to-bin fluctuations. 
As in K03 the error bars are computed with a jack-knife
estimator, which gives results very similar to a bootstrap resampling
of the data. Note that the bispectrum computed from the 
continuum fitted observed absorption spectra is negative.
The same was found by Mandelbaum et al. (2003) for a 
statistic similar to the bispectrum in their analysis of synthetic spectra
from numerical simulations. The negative sign of the bispectrum is
expected if the higher order correlations arise from gravitational
growth (Zaldarriaga et al. 2001).

For 13 of the 27 spectra we have metal line lists. 
From these lists we have produced artificial spectra which contain the 
metal lines only. The diamonds in Figure \ref{fig1} show the mean
bispectrum of these metal-line-only spectra which have a median
redshift $z=2.36$. 
For $ k\, > 0.1$ s/km the identified metal lines contribute up to 50\% to
the bispectrum.  At larger scales their contribution  is
considerably smaller.

We have also computed the flux bispectrum for the flux estimator 
F3 of K03 (see Figure 2 of K03), which is defined as 
$\delta_{F3}=\tilde F/ \langle \tilde F \rangle -1$, where $\tilde F$ 
is the flux of the spectrum without continuum fitting. 
At $k< k_{\rm cont} \sim 0.003\,$ s/km continuum fluctuations 
dominate the flux bispectrum and the values of the bispectrum 
become positive. The same was found by K03 for the flux
power-spectrum. We therefore do not plot the
bispectrum for $k<0.003$ s/km.

\subsection{Comparison with hydro-dynamical simulations}
\label{hydro}

We have also calculated the bispectrum for synthetic spectra obtained from
two outputs at $z=2.3$ from the large set of hydro-dynamical
simulations presented by Springel \& Hernquist (2003a).  The
simulations were performed with the code GADGET (Springel, Yoshida \&
White 2001) modified to conserve entropy (Springel \& Hernquist 2002)
and are for a $\Lambda$CDM model with parameters $\Omega_0=0.3$ and
$\Omega_{\Lambda}=0.7$, Hubble constant $H_0=100 \,h \vel$ Mpc$^{-1}$
with $h=0.7$, baryon density $\Omega_b=0.04$, a power spectrum with
primordial spectral index $n=1$, and {\it rms} fluctuation amplitude on
8$h^{-1}$ Mpc scale, $\sigma_8=0.9$.

The simulation  outputs are  part of  the O3 run and the D4 run. 
The O3 simulation has a box size of  
10 $h^{-1}$ Mpc box size, no feedback due to star formation 
and was performed with $2 \times 144^3$ particles.  The D4 run has a box
size of 33.75 $h^{-1}$ Mpc, $2 \times 216^3$ particles and includes 
feedback by galactic winds. The  mass of a gas particle is
$3.72\times 10^6 M_{\odot}$ and $4.24\times
10^7 M_{\odot}$ in O3 and D4, respectively. 
Both simulations  assume an UV background as modelled by Haardt 
\& Madau (1996) and follow star formation with a hybrid multi-phase model
(see Springel \& Hernquist 2003b for details). Synthetic spectra 
are extracted from the simulation in the usual way and are normalized  
to reproduce the effective optical depth $\tau_{eff}=-\ln \langle F 
\rangle=0.165$ at $\langle z \rangle=2.2$ in the observed spectra used 
for calculating the bispectrum. 

The triangles in Fig.~\ref{fig1} show the flux bispectrum of 
the synthetic spectra with and without feedback from galactic
winds.  At $k < 0.1$ s/km the difference between the two simulations 
is smaller than  $10\,\%$.  
This confirms that the feedback from high-redshift galaxies 
is expected to have  little effect on the gas responsible 
for \lya absorption (Theuns et al. 2002).  
This is most likely  because the simulations mainly differ
in the distribution of hot gas which  has a small filling
factor even for the simulation with feedback. The difference in  box 
size and resolution also appear to have a little  effect.  
For $ k<0.1$ s/km the bispectrum of synthetic and observed spectra 
agree  within the errors. At larger wavenumbers the observed
bispectrum exceeds that of the synthetic spectra. This is expected since 
at these wavenumbers the metal lines contribute significantly
to the bispectrum (see Figure \ref{fig1}).

\subsection{The bispectrum for randomized absorption spectra} 

\begin{figure}
\center\resizebox{.48\textwidth}{!}{\includegraphics{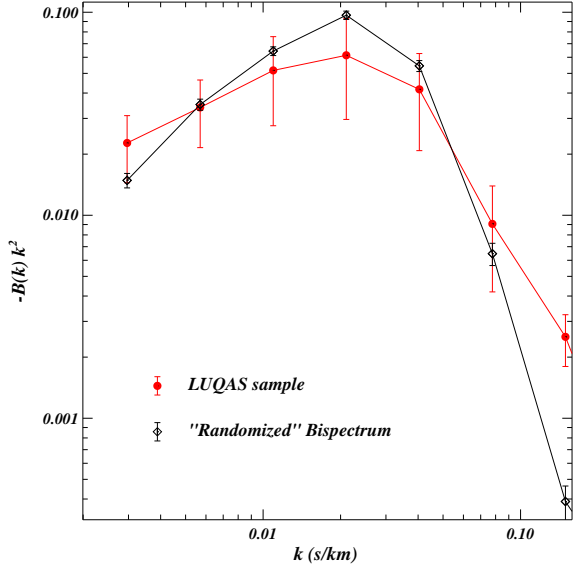}}
\caption{{\protect\footnotesize{
Comparison between the flux bispectrum of observed (filled circles) 
and  ``randomized'' absorption spectra  (empty triangles). The randomized
absorption spectra have been obtained by randomly shifting 
the position of the absorption lines as obtained with VPFIT 
within the observed wavelength range. Error bars are jack-knife 
estimates.}}}
\label{fig3}
\end{figure}

Viel et al. (2003b) have demonstrated that the  \lya forest flux 
power spectrum  of ``randomised'' QSO absorption spectra is comparable 
in shape  and amplitude to the flux power spectrum of the original
observed spectra. They found that this is because of a large 
contribution of the shape of the Voigt profiles  of discrete 
absorption systems  to the flux power spectrum. We have performed 
here a similar test for the bispectrum. For 5 of the observed spectra
in the redshift range where we have calculated the flux bispectrum we 
have a list of HI absorption lines obtained from Voigt profile fitting 
with VPFIT.  For these we have produced 50 artificial randomized
spectra each, where we have randomly shifted  the hydrogen lines in wavelength.
In Fig.~\ref{fig3} we compare the corresponding flux bispectrum
to the flux bispectrum of the observed spectra.  At  $k<0.1$
s/km they agree within the errors.  A random superposition of Voigt 
profiles appears to reproduce the observed bispectrum similarly well
as the numerical simulations.  Metal lines were omitted in the
reshuffled spectra hence the discrepancy at large wavenumbers. As in 
Viel et al. (2003) we have also calculated the bispectrum for 
randomized spectra where we  halved and doubled the Doppler parameter 
of the lines. The result is similar. Broadening and narrowing of the 
lines leads to an approximately linear shift in wavenumber. This
suggests that the bispectrum measures to a significant extent the
shape of Voigt profiles. Note that the errors of the flux bispectrum 
are  larger than that of  the flux power spectrum.  It is thus more
difficult  to discriminate between the flux bispectrum of 
randomized spectra and the flux distribution of a CDM-like model
than between the corresponding flux power spectra. 

\subsection{A second order perturbation theory approximation}
\label{theory}

The structures which give rise to the \lya forest have been 
shown to be  mildly non-linear and to be reasonably well described by 
simple approximation schemes such as the log-normal model, at least 
on large scales (Bi \& Davidsen 1997; Viel et al. 2002a, 2002b, 2003a; 
Matarrese \& Mohayaee 2002). This motivated  the use  of perturbation theory
to obtain analytical approximations in the  Fluctuating Gunn-Peterson 
Approximation to get some insight into the shape and scaling of the 
the flux power spectrum at large scales (e.g. Zaldarriaga, Scoccimarro \&
Hui 2002). In the same spirit we will here make an attempt to use 
second-order perturbation theory (e.g. Matarrese \et
1997) to derive an analytical expression  which relates flux power spectrum 
and  flux bispectrum and test it with  synthetic spectra.  

If the initial fluctuations are Gaussian and  
structures grow via gravitational instability the three-point 
correlation function is a second-order quantity. I
In the  Fluctuating Gunn-Peterson Approximation (FGPA; e.g. Hernquist
et al. 1996) the flux can be
related to the density field as 
\be
\label{fgpa}
F=\exp{[-A\,(1+\delta_{IGM})^{\beta}]}\;, 
\ee 
where $A$ and $\beta=2-0.7(\gamma-1)$ depend on $z$. $\gamma$ is the 
power-law index
of the gas temperature-density relation as usual. This assumes that
redshift space distortions are not important (but see the extra effect of this
assumption in Verde et al. 1998) and neglects thermal broadening and
instrumental noise. Moreover, we assume that at the scales of
interest $\delta_{IGM} \approx \delta_{DM} \approx \delta$.  A more
refined treatment, where these approximations are dropped, will be
presented elsewhere.

Expanding the density field to second order as $\delta ({\bf x})
\approx \delta^{(1)} ({\bf x}) + \delta^{(2)}({\bf x})$, and using the
FGPA we get 
\be \delta_F \approx b_1 [\delta^{(1)}({\bf x}) +
\delta^{(2)}({\bf x})] + \frac{b_2}{2} \delta^{(1)\,2}({\bf x})\; , 
\ee
with $b_1 = - A\, \beta$ and $b_2 = - A\, \beta\, (\beta - 1 - A\,
\beta)$. 
The second-order quantity $\delta^{(2)}$ can be expressed as a quadratic 
combination of linear perturbations $\delta^{(1)}$ 
with  a suitable convolution kernel.  The kernel is obtained by an expansion 
of the  equations of gravitational instability up to second order 
(e.g. Catelan \et 1995 and references therein).    
The expression for the \lya bispectrum follows then  by projecting the 
3D bispectrum (e.g. Matarrese \et 1997) along the line-of-sight,    
\begin{eqnarray}
\label{bisp_eq}
B(k_1,k_2,k_3)&=&\left(\frac{12}{7}\,c_1+c_2\right)p(k_1)\,p(k_2) \\
&+& c_1\,[(k_1\,k_2-\frac{2}{7}k_1^2)\,p^{(-1)}(k_1)\,p(k_2)  \nonumber \\
&+&(k_2\,k_1-\frac{2}{7}k_2^2)\,p^{(-1)}(k_2)\,p(k_1)  \nonumber \\
&+&\frac{6}{7}k_1^2\,k_2^2\,p^{(-1)}(k_1)p^{(-1)}(k_2)]+{\rm{cyc.(1,2,3)}}\;, 
\nonumber
\end{eqnarray}
where $c_1=1/b_1$, $c_2=b_2/b_1^2$. $p(k)$ is the 1D flux power
spectrum. The spectral moment $p^{(-1)}(k)$ is given by,
\be
p^{(\ell)}(k)=|k|^{2\ell}\,p(k)+2\,l \int^{\infty}_{|k|}dq 
q^{-2\ell-1}\,p(q) \;  
\label{pm1}
\ee  
with $\ell=-1$. Note that in the above eqs. $k_1,k_2,k_3$ have to
be taken including their signs. The theoretical 
relation between the flux bispectrum and power-spectrum only depends 
upon the IGM parameters $A$ and $\beta$. The very weak dependence 
on the matter density parameter $\Omega_m$ (e.g. Bouchet \et 1992; 
Catelan \et 1995) has been neglected  as usual.

\begin{figure}
\center\resizebox{.48\textwidth}{!}{\includegraphics{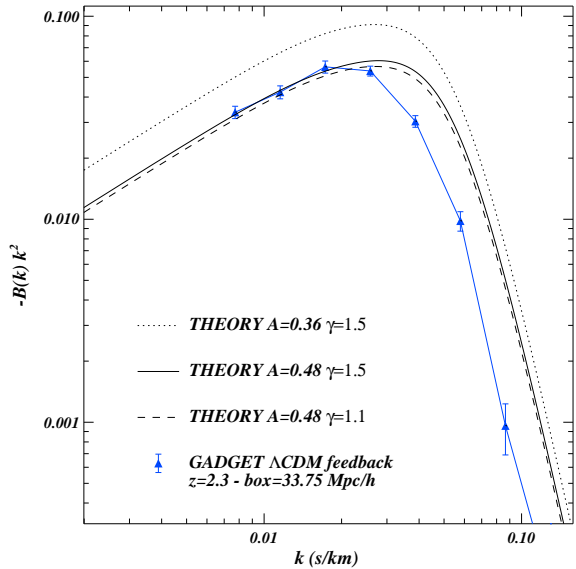}}
\caption{{\protect\footnotesize{
The flux bispectrum of the \lya forest calculated from the flux 
power spectrum  of synthetic spectra  using the analytical relation 
of eq. (\ref{bisp_eq}).  The continuous, dashed and dotted
curves are for  different parameter A and $\gamma$ 
of  the  Fluctuating Gunn-Peterson Approximation 
(see text for details). Triangles show the bispectrum 
calculated directly from the synthetic spectra.}}}
\label{fig2}
\end{figure}

In Figure \ref{fig2} we compare the bispectrum calculated 
from the 1D flux power spectrum of the sythentic spectra 
using  eqs.  (\ref{bisp_eq}) and (\ref{pm1}) with the flux bispectrum
calculated directly from the synthetic spectra.  
The reason for choosing the simulated and not the observed 
spectra to test the accuracy of the analytical relation between bispectrum 
and power-spectrum is the significantly smaller errors of the bispectrum
of the synthetic spectra.   We have used synthetic spectra 
calculated from the D4 run of the numerical simulations(see
Section \ref{hydro}).  At  $k<0.03$ s/km the relation 
between bispectrum and power spectrum based on second-order perturbation
theory is in  good agreement with the simulations 
if we choose $A=0.48$. The dashed and continuous curve 
are for $\gamma=1.1$  and $\gamma=1.5$, respectively.  
The differences between a value of $\gamma=1.1$ and $\gamma=1.5$ are very
small. At $k>0.03$ s/km the analytical prediction 
and the true bispectrum start to differ. This is probably due to
neglect of the effects of thermal broadening and Jeans smoothing.  
                    
Note that in order to reproduce the correct amplitude 
we had to choose a different value $A$ in the analytical
approximation  of the bispectrum than that used 
in the synthetic spectra where $A$  was set to reproduce the observed 
mean flux decrement.  The 
synthetic spectra were calculated with $A = 0.36$ and the
analytical prediction of the bispectrum for this value is shown by the
dotted curve in Figure \ref{fig2}. There is a clear  discrepancy in
the amplitude. We have tried to find the reason 
why the analytical prediction gives a good match for a somewhat
different value of $A$ without much success. We can only speculate 
that non-linear effects, not properly taken into account by
second-order perturbation theory, are responsible for that.

\section{Results and Discussion}
\label{conclu}
We have computed the flux bispectrum of the LUQAS, sample a set of  high
resolution quasar spectra taken with UVES in the redshift 
range $2<z<2.4$, and compared to the flux bispectrum of synthetic
spectra of a hydrodynamical simulation and that of randomized spectra. 

The main results are as follows:
\begin{itemize}
\item
at wavenumbers $k<0.1$ s/km  the flux bispectrum of the observed spectra 
and that of synthetic spectra obtained of hydro-dynamical simulations 
of a $\Lambda$CDM model agree  well  within the errors (50\%); 
\item
including feedback from galactic winds has little effect on the 
flux bispectrum; 
\item 
at wavenumbers $k>0.1$ s/km  identified metal lines 
contribute significantly to the flux bispectrum 
(of the order  50 \%), while at larger scales the 
contribution by metal lines is negligible; 
\item at wavenumbers $k<0.1$ s/km the observed flux bispectrum is consistent  
within the errors with that obtained for randomized spectra;
\item
on   scales $k < 0.03$ s/km an analytical relation  
between the flux bispectrum and flux power spectrum based  on 
second-order  perturbation theory in the framework of the  
FGPA reproduces the slope but not the amplitude 
of the bispectrum. 
\end{itemize}

In summary, the observed bispectrum obtained from the LUQAS sample agrees
similarly well with that of absorption spectra obtained from numerical
simulations and randomized observed absorption spectra. This suggests
that significantly larger samples of observed spectra as for example
expected from SDSS and  a tight control on systematic errors are
necessary to utilize the bispectrum to constrain the 3D 
distribution of the absorbers and/or cosmological parameters.

\begin{table}
\caption{Flux bispectrum$^{\rm (a)}$ of the LUQAS sample
in the redshift range $2.0<z<2.4$.}
\label{tab1}
\begin{tabular}{lc}
\hline
\noalign{\smallskip}
k (s/km) & $B(k)$ (km/s)$^{2}$ \\
\noalign{\smallskip}
\hline
0.0030 & -2845.669 $\pm$  1057.042\\
0.0057 & -1235.780 $\pm$   434.839\\
0.0109 &  -507.897 $\pm$   242.718\\
0.0210 &  -147.202 $\pm$    76.589\\
0.0404 &   -27.551 $\pm$    14.666\\
0.0777 &    -1.677 $\pm$     0.894\\
0.1494 &    -0.126 $\pm$     0.035\\
0.2872 &    -0.004 $\pm$     0.001\\
\noalign{\smallskip}
\hline
\end{tabular}
\begin{list}{}{}
\item[(a)]
Bispectrum of the quantity $\delta_F=F/<F>-1$, where  $F$ is 
the observed continuum-fitted flux and $<F>$ is the mean flux 
for each QSO, as computed from eq. (\ref{eq1}). 
\end{list}
\end{table}

\section*{Acknowledgments.} 
This work is supported by the European Community Research and
Training Network ``The Physics of the Intergalactic Medium''.
We would like to thank ESO for  making publicly available a 
superb set of QSO absorption spectra. MV thanks PPARC for financial support.

\end{document}